# EFFECT OF MASS FLUCTUATION ON COLLECTIVE MODES IN A DUSTY PLASMA

**D. Benlemdjaldi[1], R. Bharuthram[2] and R.Annou[2,3]**

[1]Department of Theoretical Physics, Faculty of Physics, USTHB, Algiers, ALGERIA

[2] School of Pure and Applied Physics, UKZN, Durban, SOUTH-AFRICA

[3]Permanent address: Department of Theoretical Physics, Faculty of Physics, USTHB, Algiers, ALGERIA

**Abstract**

The grains immersed in plasma get eroded by the ions falling onto them. As a consequence, they exhibit self-consistent mass fluctuations due to perturbations in the plasma charging currents. The grain mass is then a dynamical variable. The modifications in the plasma dielectric properties are investigated and new ultra-low frequency modes are shown to exist in the plasma.

Due to the dynamical character of the dust grain electric charge, dusty plasmas that contain dust grains of high charge and mass, exhibit new and interesting aspects with respect to already known negative-ion plasmas. Indeed, dust grains when immersed in a plasma acquire their charge via numerous and competing processes, such as photoelectric effect, thermo-ionic effect, secondary emission and plasma particles capture by the grains. As a matter of fact, the electrons are captured by the grain, whereas the ions exchange charge with it, and hence the mass of the grain do not increase appreciably. Any wave motion in the plasma induces particle density perturbations, which in turn affect self-consistently the particle currents falling onto the grains. Consequently, as a response, the grain charge would fluctuate, giving rise to many collective effects [1,2]. So far, the grain mass has been considered constant. Although, in realistic environments, e.g., astrophysical situations, the grain mass is variable. Indeed, the grain mass may grow due to particle accretion or grain-grain agglomeration. The grain may be eroded also, under the effect of numerous processes such as, sputtering that is ejection of surface atoms of the grain due to striking ions, chemical sputtering, grain-grain collisions and grain disruption [3]. The grain mass is then a dynamical variable coupled to other variables such as particle density. The effect of grain sputtering on self-gravitating dusty plasmas dielectric properties has been investigated, and it has been shown that grain mass fluctuation is at the root of a new instability [4]. In this note we propose to investigate the effect of self-consistent grain mass fluctuations that are a response to perturbations in the plasma charging currents on wave propagation along with instability processes. New modes in the ultra-low frequency range are found.

We consider a dusty plasma consisting of electrons, ions and dust grains that acquire their charge by electrons and ions attachment. During this process, dust grains are being eroded by striking ions. The governing equations are cast as follows,

$$\frac{\partial n_e}{\partial t} + \nabla \cdot n_e \vec{V}_e = -\beta_e n_e \tag{1}$$

$$\frac{\partial n_i}{\partial t} + \nabla \cdot n_i \vec{V}_i = -\beta_i n_i \tag{2}$$

$$\frac{\partial n}{\partial t} + \nabla \cdot n\vec{V} = 0 \tag{3}$$

$$m_e n_e \frac{dV_e}{dt} = -\nabla p_e - e n_e \vec{E} - m_e n_e \hat{a}_e (\vec{V}_e - \vec{V}) \tag{4}$$

$$m_i n_i \frac{dV_i}{dt} = -\nabla p_i + e n_i \vec{E} - m_i n_i \hat{a}_i (\vec{V}_i - \vec{V}) \qquad (5)$$

$$n \frac{dm\vec{V}}{dt} = qn\vec{E} - \nabla p + m_e n_e \beta_e (\vec{V}_e - \vec{V}) + m_i n_i \beta_i (\vec{V}_i - \vec{V}) \qquad (6)$$

$$\frac{dm}{dt} = -\lambda n_i m^{2/3} \qquad (7)$$

$$\frac{dq}{dt} = I_e + I_i \qquad (8)$$

$$\nabla^2 \phi = -4\pi e(n_i - n_e) - 4\pi(nq) \qquad (9)$$

$$I_e = -4\pi a^2 n_e e \sqrt{\frac{T_e}{2\pi m_e}} \exp\left(\frac{eq}{aT_e}\right) \qquad (10)$$

$$I_i = 4\pi a^2 n_i e \sqrt{\frac{T_i}{2\pi m_i}} \left(1 - \frac{eq}{aT_i}\right) \qquad (11)$$

$$m = \frac{4\pi}{3} \rho a^3, \qquad (12)$$

where, $\lambda = \sqrt{\frac{8\pi T_i}{m_i}} \left(\frac{3}{4\pi\rho}\right)^{\frac{2}{3}} m_t Y_s (1+x_0) \exp(-x_0)$, $\beta_\sigma = \frac{I_\sigma n}{q_s n_s}$, $m_t$ being the mass of the target atom constituting the grain, $\sigma$ = e,i and $Y_s$ and $x_0$ are respectively, the sputtering yield and the ratio of the binding energy of atoms of the grain surface to the thermal energy of the plasma, per unit maximum energy transfer, per unit of incident projectile energy. Equations (3) and (7) express the conservation of grains total number, and the loss of grain mass due to sputtering, where it is clear that the incident particle energy should be less than the necessary energy for grain destruction, as in this case would not be the appropriate source term. All other quantities conserve their usual meaning. It is to be stressed that the contribution of the electrons is negligible since the maximum energy transfer (for the same incident projectile energy) ratio for electrons and ions is given by $\left(\frac{m_i}{m_t} + 1\right)^2 \frac{m_e}{m_i} < 1$.

We linearize Eqs. (1-12), and the most important equation showing the effect of mass fluctuations that is the new feature developed in this note, is given by

$$\frac{d\pmb{d}q}{dt}+\pmb{h}dq=\frac{\partial I_e}{\partial n_e}dn_e+\frac{\partial I_i}{\partial n_i}dn_i-\hbar da,\tag{13}$$

where, $\pmb{h}=\left(\frac{\partial I_e}{\partial q}+\frac{\partial I_i}{\partial q}\right)_{a=cte}$, $\hbar=\left(\frac{\partial I_e}{\partial a}+\frac{\partial I_i}{\partial a}\right)_{q=cte}$ and $\pmb{d}m\propto \pmb{d}a$.

It is clear now that we have a multi-timescale problem. Indeed, around an equilibrium charge $q_0$ determined by the balance of the electric and the ionic charging currents, the grain charge $q$ fluctuates. In our work, we stress the fact that this equilibrium charge is no longer fixed, since the grain size is evolving. Consequently, the grain charge varies due to any modification of electron and ion densities, it decays also at a frequency $\pmb{h}$, and varies due to the grain radius variation as well. These causes intervene at different timescales.

Then, we assume the $\exp i(\pmb{w}-kx)$ dependence and obtain the following dispersion relation,

$$(A-A')(B''-B')=(A'-A'')(B'-B)\tag{14}$$

where, $A=\left(\overline{I}_1+\pmb{1}_1\overline{I}_4+\pmb{g}_1\overline{I}_5+\pmb{d}_1\overline{I}_6\right)/D_1$, $B=\left(\overline{I}_3+\pmb{1}_3\overline{I}_4+\pmb{g}_3\overline{I}_5+\pmb{d}_3\overline{I}_6\right)/D_1$,

$$D_1=\pmb{a}V_0-\overline{I}_2-\pmb{1}_2\overline{I}_4-\pmb{g}_2\overline{I}_5-\pmb{d}_2\overline{I}_6, A'=\left(\overline{\pmb{g}}_1-\pmb{g}_1-\overline{\pmb{g}}_4\pmb{1}_2\right)/D_2,$$

$$B'=\left(\tilde{\pmb{a}}_2-\tilde{\pmb{a}}_3-\tilde{\pmb{a}}_4\ddot{\pmb{e}}_3\right)/D_2, D_2=\pmb{g}_2-\overline{\pmb{g}}_3-\overline{\pmb{g}}_4\pmb{1}_2-\pmb{a}V_0\overline{\pmb{g}}_5,$$

$$A''=\left(\overline{\pmb{d}}_1-\pmb{d}_1+\pmb{1}_1\overline{\pmb{d}}_4\right)/D_3, B''=\left(\overline{\pmb{d}}_2-\pmb{d}_3-\pmb{1}_3\overline{\pmb{d}}_4\right)/D_3,$$

$$D_3=\pmb{d}_2-\overline{\pmb{d}}_3-\pmb{1}_2\overline{\pmb{d}}_4-\pmb{a}V_0\overline{\pmb{d}}_5, \pmb{a}=1-\frac{\pmb{w}}{kV_0}, \pmb{b}_{i0}n_{i0}=\pmb{b}_{e0}n_{e0},$$

$$\pmb{h}=\frac{|I_{e0}|}{q_0}\frac{\pmb{e}(2-\pmb{e})}{1-\pmb{e}}, \pmb{e}=\frac{eq_0}{a_0T_e}, \Delta=\frac{\pmb{y}}{3}(2-\pmb{e})$$

$$\pmb{y}=-\frac{\pmb{w}^*}{\frac{2}{3}\pmb{w}^*+i(kV_0-\pmb{w})}, \pmb{w}^*=\pmb{1}\,n_{i0}m_0^{-1/3},$$

$$\pmb{b}_1=\frac{|I_{e0}|}{\pmb{h}+i(kV_0-\pmb{w})}\left[1+\frac{\pmb{e}}{1-\pmb{e}}\Delta\right]$$

$$\pmb{b}_2=-\frac{|I_{e0}|}{\pmb{h}+i(kV_0-\pmb{w})}, \pmb{g}_1=-\frac{1}{ik}\left\{-i(kV_{e0}-\pmb{w})+\pmb{b}_{e0}\left(1-\frac{\pmb{e}\pmb{b}_2}{q_0}\right)\right\}$$

$$g_2 = -\frac{b_{e0}}{ik} \qquad g_3 = -\frac{b_{e0}}{ik}\left(\Delta + \frac{eb_1}{q_0}\right) \qquad d_1 = \frac{b_{i0}}{ik}\frac{e}{1-e}\frac{b_2}{q_0}$$

$$d_2 = -\frac{b_{i0}}{ik} \qquad d_3 = \frac{1}{ik}\left\{\frac{b_0}{1-e}\left(\frac{eb_1}{q_0}-\Delta\right)-(b_0+i(kV_{i0}-w))\right\}$$

$$\bar{g}_1 = -\frac{1}{b_{e0}-iw}\left[(V_{e0}-V_0)b_{e0}\left\{\frac{eb_2}{q_0}+1\right\}+ikv_{the}^2\right]$$

$$\bar{g}_2 = -\frac{b_{e0}(V_{e0}-V_0)}{b_{e0}-iw}\left(\Delta+\frac{eb_1}{q_0}\right) \quad ; \quad \bar{g}_3 = -\frac{b_{e0}(V_{e0}-V_0)}{b_{e0}-iw} \quad ; \quad \bar{g}_4 = \frac{-e/m_e}{(b_{e0}-iw)}$$

$$\bar{g}_5 = \frac{b_{e0}}{(b_{e0}-iw)} \quad ; \quad \bar{d}_1 = \frac{b_{e0}(V_{i0}-V_0)}{b_{i0}-iw}\frac{eb_2}{q_0(1-e)}$$

$$\bar{d}_2 = -\frac{1}{b_{i0}-iw}\left[ikn_{thi}^2+b_{i0}(V_{i0}-V_0)\left\{1+\frac{\Delta}{1-e}-\frac{b_1 e}{q_0(1-e)}\right\}\right]$$

$$\bar{d}_3 = -\frac{b_{i0}(V_{i0}-V_0)}{b_{i0}-iw} \quad ; \quad \bar{d}_4 = \frac{e/m_i}{b_{i0}-iw} \quad ; \quad \bar{d}_5 = \frac{b_{i0}}{b_{i0}-iw}$$

$$l_1 = \frac{4p}{ik}\{n_0 b_2 - en_{e0}\} \quad ; \quad l_2 = \frac{4pq_0 n_0}{ik} \quad , \quad l_3 = \frac{4p}{ik}\{en_{i0}+n_0 b_1\}$$

$$\bar{I}_1 = \left\{\frac{m_e}{m_0}\frac{n_{e0}}{n_0}(V_{e0}-V_0)b_{e0}\left(1+\frac{b_2 e}{q_0}\right)-\frac{m_i}{m_0}(V_{i0}-V_0)\frac{b_{i0} e}{1-e}\frac{b_2}{q_0}\right\}\bigg/D_4$$

$$D_4 = i(kV_0-w)+\frac{m_e}{m_0}\frac{n_{e0}}{n_0}b_{E0}+\frac{m_i}{m_0}\frac{n_{i0}}{n_0}b_{i0}$$

$$\bar{I}_2 = \left\{\frac{m_e}{m_0}\frac{n_{e0}}{n_0}(V_{e0}-V_0)b_{e0}+\frac{m_1}{m_0}\frac{n_{i0}}{n_0}(V_{i0}-V_0)b_{i0}-ikv_{th}^2\right\}\bigg/D_4$$

$$\bar{I}_3 = \left[\frac{m_i}{m_0}\frac{n_{i0}}{n_0}(V_{i0}-V_0)b_{i0}\left\{1+\frac{\Delta}{1-e}-\frac{eb_1}{q_0(1-e)}\right\}+b_{e0}\frac{m_e}{m_0}\frac{n_{e0}}{n_0}(V_{e0}-V_0)\left(\frac{eb_1}{q_0}+\Delta\right)-i(kV_0-w)V_0 y\right]\bigg/D_4$$

$$\bar{I}_4 = \frac{q_0/m_0}{D_4} \quad ; \quad \bar{I}_5 = \frac{m_e}{m_0}\frac{n_{e0}}{n_0}b_{e0}\bigg/D_4 \quad ; \quad \bar{I}_6 = \frac{m_i}{m_0}\frac{n_{i0}}{n_0}b_{i0}\bigg/D_4$$

We confine ourselves in this note to a quest of ultra-low frequency modes, as this range of frequencies, has not been investigated so far. Indeed, if one considers a non-streaming plasma ($V_{e0} = V_{i0} = V_0 = 0$), the dispersion relation (14) reduces to,

$$\gamma_1 k^4 + \gamma_2(\omega) k^2 + \gamma_3(\omega) = 0 \tag{15}$$

where, $\gamma_1 = \left(\dfrac{q_0}{e}\right)^2 \dfrac{m_0}{m_e} \dfrac{m_0}{m_i} \left(\dfrac{v_{the}}{\omega_d}\right)^2 \left(\dfrac{v_{thi}}{\omega_d}\right)^2$

$$\gamma_2 = -\dfrac{q_0}{e} \dfrac{m_0}{m_e} \left(\dfrac{v_{the}}{\omega_d}\right)^2 \left\{\gamma - 2\left(\dfrac{\ell-1}{\ell Z} - L\right)\right\}$$

$$\gamma_3 = \left(\dfrac{5}{3}L - 1\right)\gamma - 2(L-1)\left\{\dfrac{\ell-1}{\ell Z} - L\right\}$$

$$L = \dfrac{1}{Z} + \dfrac{1-\varepsilon}{\varepsilon(2-\varepsilon)} \quad ; \quad Z = \dfrac{q_0}{e} \dfrac{n_0}{n_{e0}} \quad ; \quad \ell = \dfrac{n_{e0}}{n_{i0}}$$

Distinct cases may be investigated for a constant grain mass; four modes at zero frequency are found, i.e. two oscillating modes as well as a pair of a damped and a growing mode.

$$k_1^2 = 2 \dfrac{e}{q_0} \left(\dfrac{\omega_d}{v_{thi}}\right)^2 \dfrac{m_i}{m_0}\left(L - \dfrac{\ell-1}{\ell Z}\right) \tag{16-a}$$

and

$$k_2^2 = i^2 \dfrac{m_e}{m_0} \dfrac{e}{|q_0|}\left(\dfrac{\omega_d}{v_{the}}\right)^2 (L-1)\left(L - \dfrac{\ell-1}{\ell Z}\right) \tag{16-b}$$

The corresponding wavelength is so important that no effect is seen on the length-scale related to laboratory situations; these are quasi-homogeneous profiles of the plasma. Indeed, for a hydrogen plasma of density, (c.f.Ref.[5] and references therein), $n_{i0} = 10^{13} cm^{-3}$; $n_0/n_{i0} \approx 10^{-4}$; $\varepsilon = -2.5$; $\dfrac{|q_0|}{e} \approx 1736$; $T_i \approx T_e \approx 1 eV$, one finds, $\lambda_1 = 2\pi/k_1 \approx 3R_E$ for $\rho = 0.4 g/cm^3$ and $\lambda_1 \approx 16.5 R_E$ for $\rho = 2.2 g/cm^3$; whereas $\lambda_2 = 2\pi/k_2 \approx 50 £_{M-E}$ for $\rho = 0.4 g/cm^3$ and $\lambda_2 \approx 0.7 £_{S-E}$ for $\rho = 2.2 g/cm^3$ where, $R_E$ is the earth radius, $£_{M-E}$ is the moon-earth distance and $£_{S-E}$ is the Sun-earth distance. The dependence on temperature is given by $\frac{(l_2)_2}{(l_2)_1} = \frac{T_{e2}}{T_{e1}}$. Had we considered the mass fluctuations we would have found in the ultra-low frequency case ($\omega^* \gg \omega$ viz., $\gamma \approx -\frac{3}{2}$) $l_{1,m} = 2.86 R_E$, such as $l_{1,m}/l_1 \approx 96\%$. The mass fluctuations in this range ($\omega^* \gg \omega$) reduces slightly the characteristic length. In an oxygen plasma one obtains, $l_1 = 4.56 R_E (25 R_E)$.

In the range of frequencies close to $\omega^*$, it is found that,

$$\psi = \frac{-\psi_1 k^4 - \frac{q_0}{e}\frac{m_0}{m_e}\left(\frac{v_{the}}{\omega_d}\right)^2 \left[\frac{\ell-1}{\ell Z} - L\right] k^2 + 2(L-1)\left[\frac{\ell-1}{\ell Z} - L\right]}{-\frac{q_0}{e}\frac{m_0}{m_e}\left(\frac{v_{the}}{\omega_d}\right)^2 k^2 + \frac{5L}{3} - 1} \quad (17)$$

Hence, we get a zero frequency mode such as, $\omega_r = 0$ and $\omega_i = -\omega^* \left[\frac{2}{3} + \frac{1}{\gamma}\right]$.

At long and short wavelengths, the decrement is given by, $\omega_i(k=0) \approx -\frac{\omega^*}{2}$ and $\omega_i(k=\infty) \approx -\frac{2}{3}\omega^*$. On the other hand, for $(\omega \gg \omega^*)$ but less than all the other frequencies encountered in the above-mentioned equations, a purely growing mode is excited with a growth rate that evolves from $\omega_i(0) = 0$ to $\omega_i(k=\infty) \approx 0.19 \omega^*$.

Considering streaming grains ($V_0 \neq 0$) would not introduce anything interesting, i.e., no modes are excited. However, by taking into account the ion drift velocity $(V_{i0} \neq 0)$, we show the existence of a mode of oscillation of a decrement at short wavelengths given by,

$$\omega_i/\omega^* \approx -\left[\frac{1}{Z\ell} + \frac{1-\epsilon}{\epsilon(2-\epsilon)}\right]\left(\frac{V_{i0}}{v_{thi}}\right)^2 \left(\frac{\beta_0 \ell}{v_{thi} k}\right)^2 - \frac{2}{3}, \text{ at a frequency, } \omega_r/\omega^* \approx \frac{1}{3}\left(\frac{\beta_0 \ell}{v_{thi} k}\right)\left(\frac{V_{i0}}{v_{thi}}\right).$$

To conclude, we recall that the grain charge is a dynamical variable that fluctuates in response to any density perturbations arising due to wave motion. Many collective phenomena have been indeed, attributed to this effect. In this note we show that the grain mass is a dynamical variable coupled self-consistently to other variables. We single out the sputtering effect and investigate the dielectric properties of dusty plasmas. It is shown that many new modes may be supported by the plasma that is solely due to grain mass fluctuations. It has been proved also, that new static modes are supported by a dusty plasma even though the grain mass is kept constant. It is a wiggler-type field of a characteristic length close to the earth radius as well as damped modes on a length scale of the order of the Sun-earth distance.

*Acknowledgment*:

RA acknowledges the joint financial supported of USTHB (Algeria) and UKZN (S. Africa).